\documentclass[aps,twocolumn,showpacs,byrevtex,prl,reprint,nofootinbib]{revtex4-1}
\usepackage{epsfig}
\usepackage{dcolumn}
\usepackage{bm}
\usepackage{color}
\usepackage{ulem}
\usepackage{graphics}
\graphicspath{{graphs/}}
\usepackage{subfigure}
\usepackage{overpic}		
\usepackage{times} 
\usepackage[english]{babel}
\usepackage{amsmath}
\usepackage{amssymb}
\usepackage{mathrsfs}	
\usepackage{breqn}		
\usepackage[colorlinks,linkcolor=blue,citecolor=blue]{hyperref}
\usepackage{lineno}
\addto{\captionsenglish}{%
  
   } \begin{document} \normalsize
  \parskip=5pt plus 1pt minus 1pt
\title{\boldmath First simultaneous
  measurement of $\Xi^0$ and $\bar{\Xi}^0$ asymmetry parameters in
  $\psi(3686)$ decay}

\author{
M.~Ablikim$^{1}$, M.~N.~Achasov$^{13,b}$, P.~Adlarson$^{73}$, R.~Aliberti$^{34}$, A.~Amoroso$^{72A,72C}$, M.~R.~An$^{38}$, Q.~An$^{69,56}$, Y.~Bai$^{55}$, O.~Bakina$^{35}$, I.~Balossino$^{29A}$
% ...
\\
\vspace{0.2cm}
(BESIII Collaboration)\\
\vspace{0.2cm} {\it
$^{1}$ Institute of High Energy Physics, Beijing 100049, People's Republic of China\\
% ...
$^{13}$ G.I. Budker Institute of Nuclear Physics SB RAS (BINP), Novosibirsk 630090, Russia\\
% ...
$^{29}$ INFN Sezione di Ferrara, (A)INFN Sezione di Ferrara, I-44122, Ferrara, Italy; (B)University of Ferrara,  I-44122, Ferrara, Italy\\
% ...
$^{34}$ Johannes Gutenberg University of Mainz, Johann-Joachim-Becher-Weg 45, D-55099 Mainz, Germany\\
$^{35}$ Joint Institute for Nuclear Research, 141980 Dubna, Moscow region, Russia\\
% ...
$^{38}$ Liaoning Normal University, Dalian 116029, People's Republic of China\\
% ...
$^{55}$ Southeast University, Nanjing 211100, People's Republic of China\\
$^{56}$ State Key Laboratory of Particle Detection and Electronics, Beijing 100049, Hefei 230026, People's Republic of China\\
% ...
$^{69}$ University of Science and Technology of China, Hefei 230026, People's Republic of China\\
% ...
$^{72}$ University of Turin and INFN, (A)University of Turin, I-10125, Turin, Italy; (B)University of Eastern Piedmont, I-15121, Alessandria, Italy; (C)INFN, I-10125, Turin, Italy\\
$^{73}$ Uppsala University, Box 516, SE-75120 Uppsala, Sweden\\
% ...
$^{b}$ Also at the Novosibirsk State University, Novosibirsk, 630090, Russia\\
% ...
}
}

\date{\today}
  
\begin{abstract} 
The $\Xi^0$ asymmetry parameters are measured using
entangled quantum $\Xi^0$-$\bar{\Xi}^0$ pairs from a sample of $(448.1
\pm 2.9) \times 10^6$ $\psi(3686)$ events collected with the BESIII
detector at BEPCII.  The relative phase between the transition
amplitudes of the $\Xi^0 \bar{\Xi}^0$ helicity states is measured to
be $\Delta \Phi = -0.050 \pm 0.150 \pm 0.020$~rad, which implies that
there is no obvious polarization at the current level of
statistics. The decay parameters of the $\Xi^0$ hyperon
$(\alpha_{\Xi^0}, \alpha_{\bar{\Xi}^0}, \phi_{\Xi^0}, \phi_{\bar{\Xi}^0})$ and
the angular distribution parameter $(\alpha_{\psi(3686)})$ and $\Delta
\Phi$ are measured simultaneously for the first time. In addition, the
$CP$ asymmetry observables are determined to be $A^{\Xi^0}_{CP} =
(\alpha_{\Xi^0} + \alpha_{\bar{\Xi}^0})/(\alpha_{\Xi^0} -
  \alpha_{\bar{\Xi}^0})$ $= -0.007$ $\pm$ 0.082 $\pm$ 0.025 and $\Delta
\phi^{\Xi^0}_{CP} = (\phi_{\Xi^0} + \phi_{\bar{\Xi}^0})/2$ $= -0.079$ $\pm$ 0.082 $\pm$ 0.010 rad, which are
consistent with $CP$ conservation. 
\end{abstract} 
\maketitle
\newpage


The universe began with the Big Bang, where it is commonly assumed
that matter and antimatter were created in equal amounts.  However, at
present, only traces of antimatter can be seen. $CP$ violation (CPV) is
one of the necessary conditions to possibly explain this
asymmetry~\cite{CPsource}. The existence of CPV in the decays of
$K^0$, $B^0$, and $D^0$ mesons~\cite{KCP,BCP1,BCP2,DCP}, as well as in
neutrino oscillations $\nu_l$~\cite{nuCP}, are firmly
established. However, these CPV effects are too small to explain the
large matter-antimatter asymmetry in the universe.

Recently, a technique to test CPV in the hyperon sector has been
developed by simultaneously analyzing the spin polarization and the asymmetry parameters of the entangled hyperon-antihyperon
pairs produced in the decays of the $J/\psi$, $\psi(3686)$, and
$\psi(3770)$ mesons at the BESIII experiment~\cite{xghe}. 
For cascade hyperon decays, the
angular distribution of the daughter hyperon is proportional to $(1 +
\alpha_{H}\boldsymbol{P}_H \cdot \hat{\boldsymbol{n}})$, where
$\alpha_{H}$ is the hyperon decay parameter, $\boldsymbol{P}_H$ and
$\hat{\boldsymbol{n}}$ are the hyperon polarization and the unit
vector in the direction of the daughter hyperon momentum,
respectively, both in the hyperon rest frame.  The $CP$ asymmetry is
defined as $A_{CP} = (\alpha_{H} + \alpha_{\bar{H}})/(\alpha_{H} -
\alpha_{\bar{H}})$, where the parameters $\alpha_{H}$ and
$\alpha_{\bar{H}}$ are $CP$ odd, and a nonzero $A_{CP}$ indicates
CPV.  In the Standard Model (SM), a tiny $A_{CP}$ value of $\sim
10^{-4}$~\cite{10m4} is predicted in the hyperon sector.  Therefore,
a test of CPV in hyperon decays is sensitive to possible sources of CPV
from physics beyond the SM~\cite{theory, Liu:2023xhg}. At present, BESIII has
performed CPV tests in the decays of
$\Lambda$~\cite{Lambda,3773Lambda,zhangjianyu},
$\Sigma^+$~\cite{Sigmap}, and $\Xi^-$~\cite{Patrik,zhangjingxu}
hyperons, where for $\Xi^-$ hyperons, the most precise asymmetry
parameter measurement was reported in $J/\psi$ decay.  BESIII has also
performed the first determination of the weak phase of the $\Xi^-$ hyperon
using entangled $\Xi^-\bar\Xi^+$ pairs~\cite{Patrik}.  However,
CPV in $\Xi^0$ hyperon decays has not so far been searched for,
the asymmetry parameter $\alpha_{\Xi^{0}}$ in the decay $\Xi^0 \to
\Lambda \pi^0$ has not been measured directly, only the product
$\alpha_{\Lambda} \cdot \alpha_{\Xi^{0}}$ has been
reported~\cite{product1,product2}, and the weak decay phase
$\phi_{\Xi^0}$ was measured with large uncertainty~\cite{HBC1,HBC2,HBC3}. 

The process of $e^+ e^- \to \psi(3686) \to \Xi^0 \bar{\Xi}^0 \to \pi^0
\pi^0 \Lambda \bar{\Lambda}$ with $\Lambda \to p \pi^-$ can be fully
described by the vector $\boldsymbol{\xi} =
(\theta_\Xi,\theta_\Lambda,\varphi_\Lambda,\theta_{\bar{\Lambda}},\varphi_{\bar{\Lambda}},\theta_{p},\varphi_{p},\theta_{\bar{p}},\varphi_{\bar{p}})$,
where the coordinate systems and angles are shown in Fig.~\ref{xyz}
with the same convention as Refs.~\cite{Patrik} and
\cite{zhangjingxu}. The cascade $\Xi$ and $\Lambda$ polarization
vector, $\boldsymbol{P}_{\Xi}$ and $\boldsymbol{P}_{\Lambda}$, are
related as $\boldsymbol{P}_{\Lambda} = \alpha_{\Xi^0}
\hat{\boldsymbol{z}}_{\Xi^0} + \beta_{\Xi^0} (\boldsymbol{P}_{\Xi^0}
\times \hat{\boldsymbol{z}}_{\Xi^0}) + \gamma_{\Xi^0}
       [\hat{\boldsymbol{z}}_{\Xi^0} \times (\boldsymbol{P}_{\Xi^0}
         \times \hat{\boldsymbol{z}}_{\Xi^0})]$, where
       $\alpha_{\Xi^0}$, $\beta_{\Xi^0}$, and $\gamma_{\Xi^0}$ are
       defined in Ref.~\cite{LeeYang} and
       $\hat{\boldsymbol{z}}_{\Xi^0}$ is the unit vector in the
       direction of the $\Xi$ momentum.  The joint angular
       distribution function is described by~\cite{formula}

\begin{dmath}
	\label{Formula}
	\mathscr{W} = \mathscr{W}(\boldsymbol{\xi},\boldsymbol{\Omega}) = \sum^{3}_{\mu,\bar{\nu}=0} \sum^{3}_{\mu'=0} \sum^{3}_{\bar{\nu}'=0} C_{\mu \bar{\nu}} a^{\Xi}_{\mu \mu'} a^{\Lambda}_{\mu' 0} a^{\bar{\Xi}}_{\nu \bar{\nu}'} a^{\bar{\Lambda}}_{\bar{\nu}' 0},
\end{dmath}

\noindent where $C_{\mu \nu}$ is the
production spin density matrix, $a_{\mu\nu}$ is the joint decay
amplitude, and $\boldsymbol{\Omega} = (\alpha_{\psi(3686)}, \Delta \Phi,
\alpha_{\Xi^0}, \phi_{\Xi^0},\alpha_{\Lambda}, \alpha_{\bar{\Lambda}},
\alpha_{\bar{\Xi}^0}, \phi_{\bar{\Xi}^0})$ is the set of decay
parameters.  The definitions of $C_{\mu \nu}$ and $a_{\mu \nu}$ may be
found in Ref.~\cite{formula}.

\begin{figure}[!htp]
	\centering
	\includegraphics[scale=0.4
]{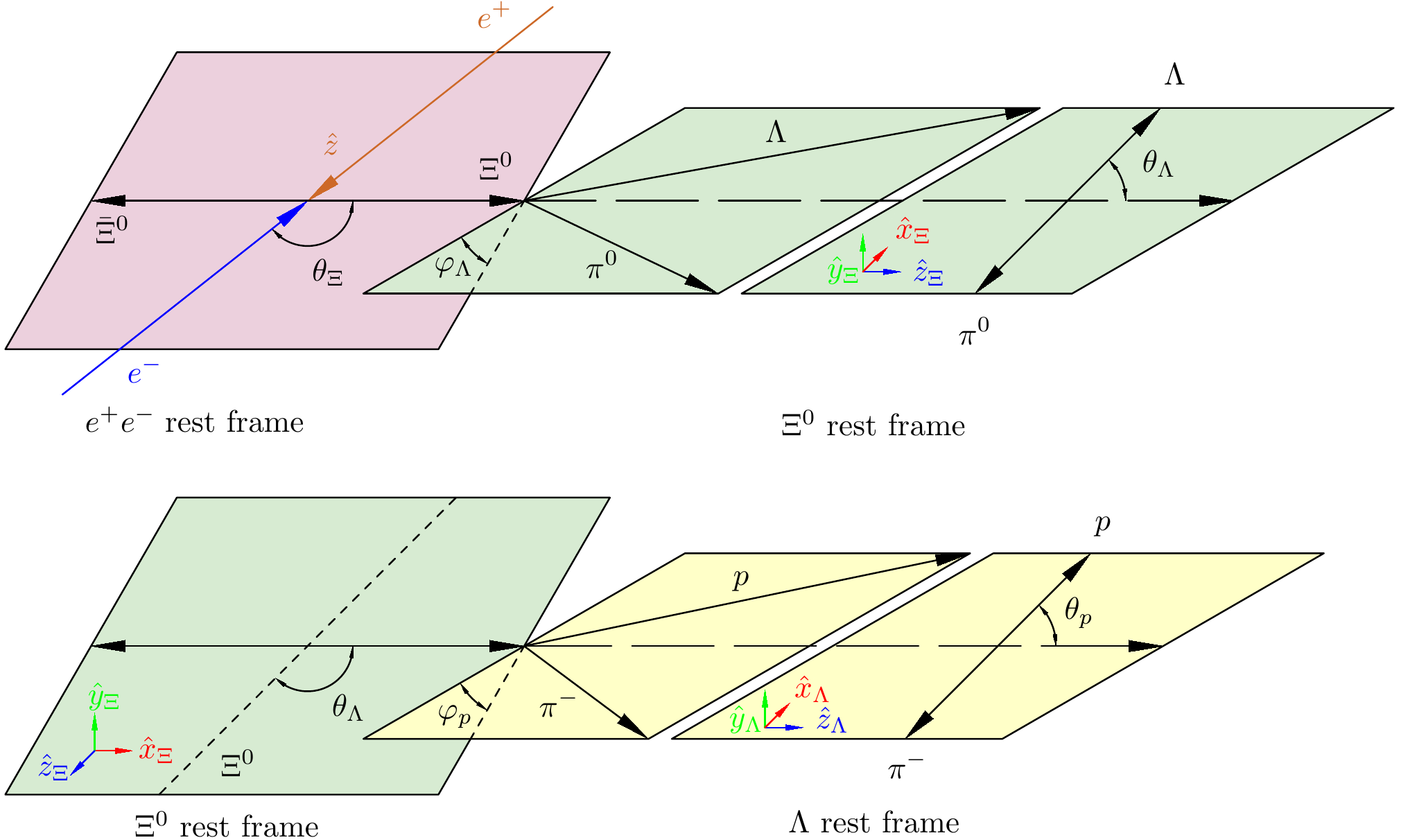}
	\caption{Depiction of the axes orientation used in the analysis of the $\Xi^0$ decay parameters. In the $e^+e^-$
          rest frame, the $\hat{\boldsymbol{z}}$ axis is
          along the $e^+$ direction, and
          $\hat{\boldsymbol{z}}_{\Xi^0}$ is along the $\Xi^0$ momentum
          direction. In the $\Xi^0$ rest frame, the polar axis
          direction is $\hat{\boldsymbol{z}}_{\Xi^0}$,
          $\hat{\boldsymbol{y}}_{\Xi^0}$ is along
          $\hat{\boldsymbol{z}} \times \hat{\boldsymbol{z}}_{\Xi^0}$
          and $\hat{\boldsymbol{z}}_{\Lambda}$ is along the $\Lambda$
          momentum direction. For the $\Lambda$ rest frame,
          the polar axis direction is
          $\hat{\boldsymbol{z}}_{\Lambda}$ and
          $\hat{\boldsymbol{y}}_{\Lambda}$ is along
          $\hat{\boldsymbol{z}}_{\Xi^0} \times
          \hat{\boldsymbol{z}}_{\Lambda}$. The vector
          $\hat{\boldsymbol{P}}_{\Xi^0} \times
          \hat{\boldsymbol{z}}_{\Lambda}$ is along the
          $\hat{\boldsymbol{y}}_{\Lambda}$ axis. The definition for the $\bar{\Xi}^0$ is analogous, with the $\hat{\boldsymbol{z}}_{\bar{\Xi}^0}$ axis against the $\hat{\boldsymbol{z}}_{\Xi^0}$ direction.}
	\label{xyz.pdf}
\end{figure}

\noindent CPV is searched
for with an amplitude, $A_{CP}$, and a phase, $\Delta \phi_{CP}$, defined as

\begin{dmath}
\label{APhi$CP$}
	\begin{aligned}
		A_{CP}^{\Xi^0} &= \frac{\alpha_{\Xi^0} + \alpha_{\bar{\Xi}^0}}{\alpha_{\Xi^0} - \alpha_{\bar{\Xi}^0}},  \\
		\Delta \phi_{CP}^{\Xi^0} &= \frac{1}{2} \left( \phi_{\Xi^0} + \phi_{\bar{\Xi}^0} \right).
	\end{aligned}
\end{dmath}

\noindent The polarization observable $P_y$
is defined as follows~\cite{polarization},
\begin{dmath}
    \label{mu}
     P_y=\frac{\sqrt{1 - \alpha_{\psi(3686)}^2 } \sin 2\theta_\Xi \sin \Delta \Phi}{2(1 + \alpha_{\psi(3686)} \cos^2 \theta_\Xi)},
\end{dmath}

\noindent which is dependent on the transverse polarization parameter $\Delta \Phi$.

In this paper, we present the first simultaneous measurement of the
$\Xi^0$ asymmetry parameters using entangled $\Xi^0$-$\bar\Xi^0$ pairs
from $(448.1 \pm 2.9) \times 10^6$ $\psi(3686)$ decays~\cite{psip}
collected with the BESIII detector~\cite{BESIII}. In addition, a study
of the transverse polarization in $\psi(3686) \to \Xi^0 \bar{\Xi}^0$ and a test of CPV in $\Xi^0$ hyperon
decays are performed.

Candidate $\psi(3686) \to \Xi^0 \bar{\Xi}^0$ events are selected
by fully reconstructing the subsequent decays $\Xi^0 \to \pi^0
\Lambda$, $\Lambda \to p \pi^-$ and $\pi^0 \to \gamma \gamma$ (as well
as the charge conjugate final states for $\bar{\Xi}$ and
$\bar{\Lambda}$ decays).  Potential background contributions are
studied with an inclusive Monte Carlo (MC) simulation sample of $\psi(3686)$
decays~\cite{TopoAna}, and an exclusive simulation of the signal
process with $5\times10^6$ events is generated with a phase space
model for normalization. The production of the $\psi(3686)$ resonance
for both MC samples is simulated with the \textsc{kkmc}
generator~\cite{kkmc1,kkmc2}, and the subsequent decays are processed
by \textsc{evtgen}~\cite{evtgen1, evtgen2}. Additionally for the
inclusive MC sample, the branching fractions of cascade decays are fixed
according to the Particle Data Group
(PDG)~\cite{PDG2022}. All the remaining unmeasured decay modes are
generated with \textsc{lundcharm}~\cite{lundcharm1,lundcharm2}.  The
response of the BESIII detector is modeled with MC simulations using a
framework based on \textsc{geant}{\footnotesize 4}~\cite{geant41,
geant42}.

Candidate events are required to contain at least four charged
particles (two positive and two negative) and at least four
photons. Charged particles are reconstructed as tracks within the
multilayer drift chamber (MDC). Only tracks fully contained in the
acceptance region of the MDC, $|\cos\theta| < 0.93$ ({with $\theta$
defined with respect to the $z$-axis, which is the symmetry axis of
the MDC}), are kept for the analysis.
Because of the momentum separation in the two body decay, the momenta
of (anti-)proton and charged pion candidates are required to be
greater and less than 0.5 GeV/$c$, respectively.

$\Lambda$($\bar\Lambda$) candidates are reconstructed as $p\pi^-$
($\bar{p}\pi^+$) pairs that satisfy a vertex fit. The four-track
combination that minimizes $\sqrt{(M_{p\pi^{-}}-m_{\Lambda})^{2} +
(M_{\bar{p}\pi^{+}}-m_{\Lambda})^{2}}$ is selected, where
$M_{p\pi^{-}}(M_{\bar{p}\pi^{+}})$ is the invariant mass of the
$p\pi^{-}(\bar{p}\pi^{+})$ pair and $m_\Lambda$ is the $\Lambda$ mass
from the PDG~\cite{PDG2022}.  To further suppress non-$\Lambda$
background, the $\Lambda$ decay length is required to be greater than
zero, where negative decay lengths are caused by the detector
resolution and background contributions.

Photon candidates are reconstructed from isolated showers in the electromagnetic calorimeter (EMC). The energy deposited in the nearby time of flight (TOF) counter is included to improve the reconstruction efficiency and energy resolution. 
The shower energies are required to be greater than 25 MeV in the EMC barrel region ($|\cos\theta|<0.8$), or greater than 50 MeV in the EMC end-cap region ($0.86 < |\cos\theta| < 0.92$). 
In order to reject electronic noise and energy deposits unrelated to the event start time, the EMC shower time, measured with respect to the collision signal, is required to satisfy $0 < t < 700$ ns.

To further suppress background from soft $\pi^{0}$s and radiated
photon events and to improve the mass resolution, a six-constraint
(6C) kinematic fit is applied to all possible
$\gamma\gamma\gamma\gamma\Lambda\bar{\Lambda}$ combinations by imposing
energy-momentum conservation and constraining the masses of the two
pairs of photons from the $\pi^0$ mesons to the $\pi^0$ mass. The
$\Xi^0$ and $\bar{\Xi}^0$ candidates are then reconstructed as the
$\pi^0 \Lambda$ and $\pi^0 \bar{\Lambda}$ combinations that minimize
the discriminant $\delta = \sqrt{(M_{\pi^0\Lambda} - m_{\Xi^0})^2 +
(M_{\pi^0 \bar{\Lambda}} - m_{\Xi^0})^2}$ from all
$\pi^0\Lambda(\pi^0\bar\Lambda)$ combinations, where $M_{\pi^0
\Lambda}(M_{\pi^0 \bar{\Lambda}})$ is the invariant mass of the
$\pi^0\Lambda(\pi^0\bar\Lambda)$ system and $m_{\Xi^0}$ is the $\Xi^0$
mass from the PDG~\cite{PDG2022}.  Finally, background contributions
from the $\psi(3686) \to \pi^0\pi^0 J/\psi$ process are rejected by
requiring the recoil mass of $\pi^0\pi^0$ combinations to be at least
$20$ MeV/$c^2$ away from the nominal $J/\psi$
mass~\cite{PDG2022}. Figure~\ref{box} shows the distribution of
$M_{\pi^0 \Lambda}$ versus $M_{\pi^0 \bar{\Lambda}}$ for candidate
events selected in data.  A clear signal around the
$\Xi^0(\bar{\Xi}^0)$ mass is observed.  Signal events are required
to simultaneously satisfy $|M_{\pi^0\Lambda}-m_{\Xi^0}| < 15$
MeV/$c^2$ and $|M_{\pi^0 \bar{\Lambda}}-m_{\Xi^0}| < 15$ MeV/$c^2$
(region marked as $\rm{S}$ in Fig.~\ref{box}).  Most of the background
contributions arise from $\psi(3686)$ decays that do not contain a
$\Xi^0\bar{\Xi}^{0}$ pair, such as $\psi(3686) \to \pi^0\pi^0 \Lambda
\bar{\Lambda}$.  The background yield is evaluated by the mean of the
three sideband regions ${\rm B}_{i}$ (with $i=1,2,3$) depicted in
Fig.~\ref{box}.  The sideband regions have the same size as the signal
region, but are centered on the following values of $(M_{\pi^0
\Lambda},M_{\pi^0 \bar{\Lambda}})=(1.27,1.27)$, $(1.36,1.27)$, and $(1.27,1.36)$ GeV/$c^2$. The sideband region
$\rm{B}_4$ is not suitable for background evaluation as it is close to
the $\psi(3686) \to \Sigma(1385)^0 \bar{\Sigma}(1385)^0$ region and
would lead to an overestimation of the signal contamination.  In the
signal region $N=1934$ events are counted with an expected background
contribution of $23 \pm 5$ events, resulting in a $1.2$\%
contamination level. The signal contamination can, therefore, be
considered as negligible in the following analysis.

\begin{figure}[!htp]
\centering \includegraphics[scale=0.445]{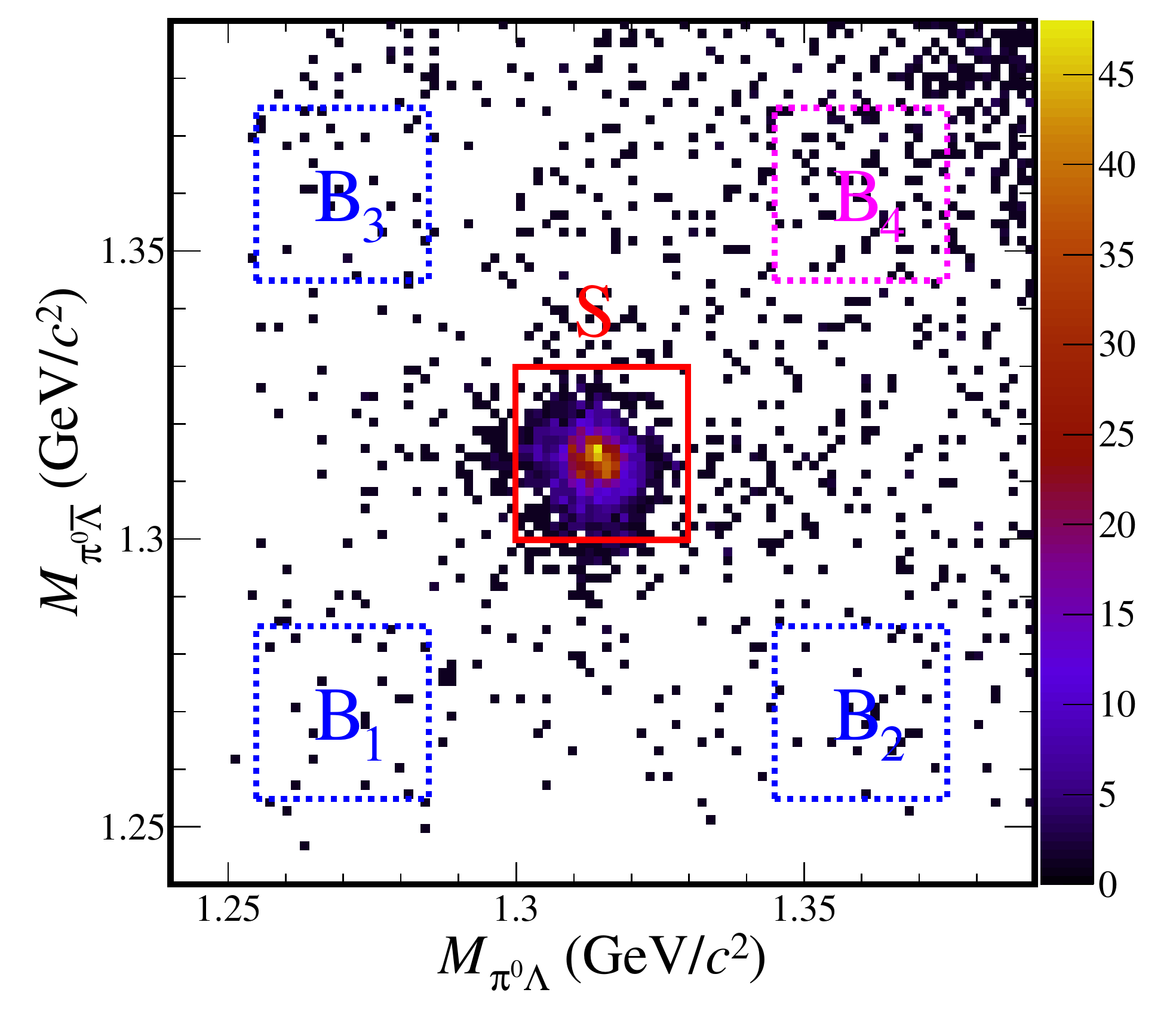} \caption{The scatter
plot of $M_{\pi^0 \Lambda}$ versus $M_{\pi^0 \bar{\Lambda}}$ of the
candidate events selected from data, where the red box S shows the
signal region, the blue boxes ${\rm B_1}$, ${\rm B_2}$, and ${\rm B_3}$
denote the selected sideband regions, and the magenta box ${\rm B_4}$
is close to the $\Sigma(1385)^0
\bar{\Sigma}(1385)^0$ signal and is not used.}  \label{box.pdf}
\end{figure} 

To determine the set of $\boldsymbol{\Omega}$ parameters, an unbinned maximum likelihood fit
(MLL fit) is performed, where the decay parameters $\alpha_\Lambda$
and $\alpha_{\bar{\Lambda}}$ are fixed to 0.754 \cite{Lambda} assuming
$CP$ conservation in $\Lambda$ and $\bar{\Lambda}$ decays.  In the
fit, the likelihood function $\mathscr{L}$ is given by

\begin{dmath}
    \mathscr{L} = \prod^N_{i=1} \frac{\mathscr{W}(\boldsymbol{\xi}_i, \boldsymbol{\Omega})\epsilon(\boldsymbol{\xi}_i)}{\mathscr{N}(\boldsymbol{\Omega})},
\end{dmath}

\noindent where the joint angular distribution $\mathscr{W}$ is
defined in Eq.~(\ref{Formula}), $N$ is the number of data events,
$\epsilon = N_{\rm survive}^{\rm MC}/N_{\rm total}^{\rm MC}$ is the
detection efficiency, and $\mathscr{N}(\boldsymbol{\Omega}) = \int
\mathscr{W}(\boldsymbol{\xi}_i, \boldsymbol{\Omega})
\epsilon(\boldsymbol{\xi}_i) \text{d}\boldsymbol{\xi}_i$ is the
normalization factor. Since the low background level has a negligible
effect, we do not include a background term in the fit, and the
parameters are determined by minimizing the function $S = -\ln
\mathscr{L}$. The fit results are reported in Table~\ref{numerical}.

\begin{table}[!htp]
	\centering
	\caption{Numerical results of parameters, where the first uncertainty is statistical and the second is systematic.}
\scalebox{0.85}{
\begin{tabular}{cr@{.}lr@{.}lr@{.}l}
\hline \hline
Param.				&\multicolumn{2}{c}{This work} 	&\multicolumn{2}{c}{BESIII~\cite{preXi0}}					&\multicolumn{2}{c}{PDG~\cite{PDG2022}} \\
\hline
$\alpha_{\psi(3686)}$	&$0$&$665 \pm 0.086 \pm 0.081$	&$0$&$650 \pm 0.090 \pm 0.140$		&\multicolumn{2}{c}{$\cdots$} \\
$\Delta \Phi$			&$-$0&$050 \pm 0.150 \pm 0.020$ 	&\multicolumn{2}{c}{$\cdots$} 					&\multicolumn{2}{c}{$\cdots$} \\
$\alpha_{\Xi^0}$			&$-0$&$358 \pm 0.042 \pm 0.013$ 	&\multicolumn{2}{c}{$\cdots$}					&$-0$&$356 \pm 0.011$ \\
$\phi_{\Xi^0}$			&$0$&$027 \pm 0.117 \pm 0.011$ 	&\multicolumn{2}{c}{$\cdots$}					&$0$&$366 \pm 0.209$ \\
$\alpha_{\bar{\Xi}^0}$	&$0$&$363 \pm 0.042 \pm 0.013$	&\multicolumn{2}{c}{$\cdots$}					&\multicolumn{2}{c}{$\cdots$} \\
$\phi_{\bar{\Xi}^0}$		&$-0$&$185 \pm 0.116 \pm 0.017$ 	&\multicolumn{2}{c}{$\cdots$}					&\multicolumn{2}{c}{$\cdots$} \\
\hline
$A_{CP}^{\Xi}$			&$-0$&$007 \pm 0.082 \pm 0.025$	&\multicolumn{2}{c}{$\cdots$}					&\multicolumn{2}{c}{$\cdots$} \\
$\Delta \phi_{CP}^{\Xi}$		& $-0$&$079 \pm 0.082 \pm 0.010$	&\multicolumn{2}{c}{$\cdots$}					&\multicolumn{2}{c}{$-$} \\
\hline \hline
	\end{tabular}
}
	\label{numerical}
\end{table}

Systematic uncertainties arise from the difference of detection
efficiencies between data and simulations (tracking, $\pi^0$ and
$\Xi^0(\bar{\Xi}^0)$ reconstruction, 6C kinematic fit, $\pi^0\pi^0
J/\psi$ background veto) as well as from the sideband technique,
background from the continuum process $e^+e^- \to \Xi^0 \bar{\Xi}^0$,
the uncertainties of the $\Lambda \to p\pi^-$ decay parameters and the
MLL fit method as listed in Table~\ref{SU}. Correction factors for tracking and $\pi^0$ reconstruction efficiency differences between data and simulations are evaluated on a control sample of $\psi(3686) \to \Xi^0 \bar{\Xi}^0$ events, where one of the hyperon is fully reconstructed and one of the charged particles ($p$, $\bar{p}$, $\pi^+$ and $\pi^-$) or the $\pi^0$ from the second is not considered. 
The uncertainties arising from the correction procedure are evaluated by mean of 100 variation of the correction factors, following a Gaussian distribution with the nominal value as mean and the statistical uncertainty as width. The difference between the nominal results of the decay parameters and the mean values of those obtained through the variations is regarded as systematic uncertainty.  The correction to the $\Xi^0$reconstruction
efficiency differences is evaluated as in Ref.~\cite{preXi0,
  BESIII:2016ssr, BESIII:2019dve, BESIII:2019cuv, BESIII:2021aer,
  BESIII:2020ktn, BESIII:2021gca, BESIII:2021ccp, BESIII:2022mfx}, and
the same procedure is applied to estimate the systematic uncertainty.
The 6C kinematic fit is sensitive to differences in the momentum
resolution of the charged tracks between data and simulations.
Corrections to the helix parameters of charged tracks are evaluated
and applied in the measurement, and the difference between the spin
polarization parameters obtained with and without the corrections is
considered as the systematic uncertainty.  The systematic uncertainty
from vetoing $\psi(3686) \to \pi^0 \pi^0 J/\psi$ background is
estimated by varying the range of the mass window requirement by 5
MeV/$c^2$. The largest difference is taken as the uncertainty.  The
uncertainties related to background contributions (sideband evaluation
and continuum process) are evaluated by introducing a background term
to the MLL function $S \to S' = -\ln \mathscr{L} + \ln
\mathscr{L}_{\rm{BKG}}$. The difference in the fit results is taken as
the uncertainty.  The possible bias introduced by the $\Lambda \to
p\pi^-$ decay parameters are estimated by changing $a_\Lambda$ and
$a_{\bar{\Lambda}}$ values reported in Ref.~\cite{Lambda} by $\pm 1
\sigma$. The largest variation with respect to the central value is
considered as the systematic uncertainty. To validate the fit
procedures, an input-output check based on 300 pseudo-experiments is
performed with the helicity amplitude formula Eq.~(\ref{Formula}). The
polarization and the asymmetry decay parameters measured in this
analysis are used as input in the formula. The number of events in
each generated MC sample is 5000, and the check is performed
independently 300 times.  The difference between the input and output
Gaussian fit values is taken as the systematic uncertainty caused by
the fit method.  Assuming all sources to be independent, the total
systematic uncertainties in the measurement of $\alpha_{\bar{\Xi}^0}$,
$\Delta \Phi$ and the decay asymmetry parameters via analyzing for
$\psi(3686) \to \Xi^0 \bar{\Xi}^0$ are determined as the sum in
quadrature of the mentioned sources.
\begin{table}[!htp]
\centering
\caption{Systematic uncertainties of the measured parameters.}
    \scalebox{0.9}{
	\begin{tabular}{lcccccccc}
\hline \hline
Source &$\alpha_{\psi(3686)}$	&$\Delta \Phi$	&$\alpha_{\Xi^0}$
&$\phi_{\Xi^0}$	        &$\alpha_{\bar{\Xi}^0}$
&$\phi_{\bar{\Xi}^0}$ \\
\hline
Tracking efficiency	    &$0.067$		&$0.003$		&$0.003$		            &$0.004$	    &$0.002$		   &$0.004$ \\
$\pi^0$ reconstruction  &$0.032$		&$0.001$		&$0.001$		            &$0.002$	    &$0.001$		&$0.000$ \\
$\Xi^0$ reconstruction	&$0.024$		&$0.003$		&$0.001$		            &$0.001$	    &$0.000$		&$0.001$ \\
6C kinematic fit		&$0.006$		&$0.003$		&$0.001$		            &$0.001$	    &$0.000$		&$0.001$ \\
$\pi^0 \pi^0 J/\psi$ background veto		&$0.016$		&$0.017$		            &$0.009$		&$0.004$	    &$0.007$		            &$0.013$ \\
Sideband subtraction		&$0.011$		&$0.000$		&$0.004$		            &$0.006$	    &$0.002$		&$0.002$ \\
Continuum process			&$0.011$		&$0.000$		&$0.004$		            &$0.006$	    &$0.002$		&$0.003$ \\
$\Lambda$ decay parameter				&$0.001$		&$0.006$		            &$0.006$		&$0.004$	    &$0.005$		            &$0.000$ \\
Fit method 									&$0.008$		        &$0.009$		&$0.003$		&$0.002$	    &$0.008$        &$0.010$ \\
\hline
Total 										&$0.081$		        &$0.020$		&$0.013$		&$0.011$	    &$0.013$		        &$0.017$ \\
\hline \hline
	\end{tabular}}
\label{SU}
\end{table}

\begin{figure}
    \centering
\includegraphics[scale=0.445]{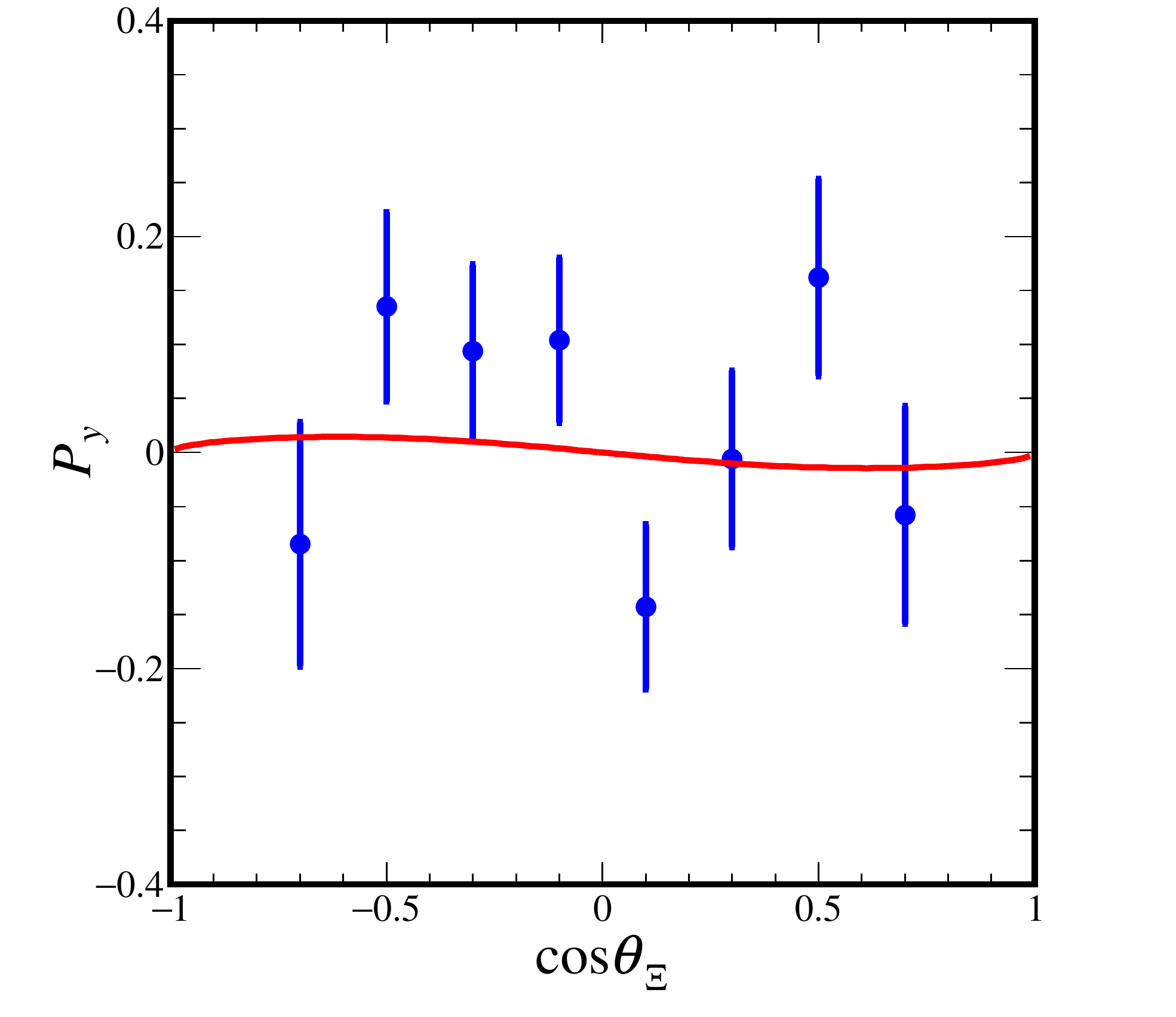}
    \caption{Distribution of the polarization observable $P_y$ versus
      $\cos \theta_{\Xi}$, dots with error bars represent experimental
      data, and the red line denotes the global fit result.}
    \label{mu1}
\end{figure}

In summary, based on a data sample of $(448.1 \pm 2.9) \times 10^6$
$\psi(3686)$ events collected with the BESIII detector, the $\Xi^0$
asymmetry parameters are measured with high precision using entangled
quantum $\Xi^0$-$\bar{\Xi}^0$ pairs.  The numerical results are
summarized in Table~\ref{numerical}.  The polarization signal related
with Eq.~(\ref{mu}) is shown in Fig.~\ref{mu1}.  The value of
$\alpha_{\psi(3686)}$ is measured to be $0.665 \pm 0.086 \pm 0.081$,
which is consistent with the previous BESIII measurement
\cite{preXi0}, and $\Delta \Phi = -0.050 \pm 0.150 \pm 0.020$~rad is
measured for the first time and is significantly different from the
one for $\Xi^-$ reported in Refs.~\cite{Patrik,zhangjingxu}.  The
relative phase that is approximately zero implies an insignificant
transverse polarization, which differs from the polarization observed
in $\Lambda$ decays from $J/\psi$ decays, $\Sigma^{+}$ decays from
both $J/\psi$ and $\psi(3686)$ decays, and $\Xi^-$ decays from both
$J/\psi$ and $\psi(3686)$
decays~\cite{Lambda,Sigmap,Patrik,zhangjingxu}. The asymmetry
parameters $\alpha_{\Xi}$ and $\alpha_{\bar{\Xi}}$ are determined
simultaneously for the first time.  Previously only the product of
$\alpha_{\Lambda} \cdot \alpha_{\Xi^0}$~\cite{product1,product2} was
reported. The parameter $\phi_{\Xi^0}$ is measured more precisely
compared with the value reported by the HBC group almost half a
century ago~\cite{HBC1,HBC2,HBC3}.  In addition, the $\Xi^0$ hyperon
$CP$ asymmetry parameters $A_{CP}^\Xi$, $\Delta\phi_{CP}^\Xi$, as
summarized in Table \ref{numerical}, indicate no CPV effect at the
current level of accuracy.  It is expected that the test of CPV will
reach sensitivities comparable to the SM prediction when a large data
sample will be available at BESIII~\cite{newdata}, the upcoming PANDA
experiment at FAIR~\cite{panda}, and the proposed Super Tau-Charm
Factory projects in China and Russia~\cite{superTC,STCF}.

The BESIII Collaboration thanks the staff of BEPCII and the IHEP computing center for their strong support. This work is supported in part by National Key R\&D Program of China under Contracts Nos. 2020YFA0406400, 2020YFA0406300; National Natural Science Foundation of China (NSFC) under Contracts Nos. 
12075107, 12247101, 11635010, 11735014, 11835012, 11905236, 11935015, 11935016, 11935018, 11961141012, 12022510, 12025502, 12035009, 12035013, 12047501, 12061131003, 12192260, 12192261, 12192262, 12192263, 12192264, 12192265, 12225509; the Chinese Academy of Sciences (CAS) Large-Scale Scientific Facility Program; the CAS Center for Excellence in Particle Physics (CCEPP); Joint Large-Scale Scientific Facility Funds of the NSFC and CAS under Contract No. U1832207; CAS Key Research Program of Frontier Sciences under Contracts Nos. QYZDJ-SSW-SLH003, QYZDJ-SSW-SLH040; 100 Talents Program of CAS; The Institute of Nuclear and Particle Physics (INPAC) and Shanghai Key Laboratory for Particle Physics and Cosmology; ERC under Contract No. 758462; European Union's Horizon 2020 research and innovation programme under Marie Sklodowska-Curie grant agreement under Contract No. 894790; German Research Foundation DFG under Contracts Nos. 443159800, 455635585, Collaborative Research Center CRC 1044, FOR5327, GRK 2149; Istituto Nazionale di Fisica Nucleare, Italy; Ministry of Development of Turkey under Contract No. DPT2006K-120470; National Research Foundation of Korea under Contract No. NRF-2022R1A2C1092335; National Science and Technology fund; National Science Research and Innovation Fund (NSRF) via the Program Management Unit for Human Resources \& Institutional Development, Research and Innovation under Contract No. B16F640076; Polish National Science Centre under Contract No. 2019/35/O/ST2/02907; Suranaree University of Technology (SUT), Thailand Science Research and Innovation (TSRI), and National Science Research and Innovation Fund (NSRF) under Contract No. 160355; The Royal Society, UK under Contract No. DH160214; The Swedish Research Council; U. S. Department of Energy under Contract No. DE-FG02-05ER41374.


\begin{thebibliography}{99}
\bibitem{CPsource} A. D. Sakharov, \href{https://doi.org/10.1070/PU1991v034n05ABEH002497}{Psi'ma Zh. Eksp. Teor.~Fiz. {\bf 5}, 32 (1967).}
\bibitem{KCP} J. H. Christenson, J. W. Cronin, V. L. Fitch, and R. Turlay, \href{https://doi.org/10.1103/PhysRevLett.13.138}{Phys. Rev. Lett. {\bf 13}, 138 (1964).}
\bibitem{BCP1} B. Aubert {\it et al.} ({\it BABAR} Collaboration), \href{https://doi.org/10.1103/PhysRevLett.87.091801}{Phys. Rev. Lett { \bf 87}, 091801 (2001).}
\bibitem{BCP2} K. Abe {\it et al.} (Belle Collaboration), \href{https://doi.org/10.1103/PhysRevLett.87.091802}{Phys. Rev. Lett. {\bf 87}, 091802 (2001).}
\bibitem{DCP} R. Aaij {\it et al.} (LHCb Collaboration), \href{https://doi.org/10.1103/PhysRevLett.122.211803}{Phys. Rev. Lett. {\bf 122}, 211803 (2019).}
\bibitem{nuCP} K. Abe {\it et al.} (T2K Collaboration), \href{https://doi.org/10.1038/s41586-020-2177-0}{Nature {\bf 580}, 339 (2020).}
\bibitem{xghe}
X.~G.~He, J.~Tandean and G.~Valencia, \href{https://doi.org/10.1016/j.scib.2022.08.012}{Sci. Bull. \textbf{67}, 1840 (2022).}
\bibitem{10m4} J. F. Donoghue, X. G. He and S. Pakvasa, \href{https://doi.org/10.1103/PhysRevD.34.833}{Phys. Rev. D {\bf 34}, 833 (1986).}
\bibitem{theory} D. G. Ireland, M.D\"oring, D. I. Glazier, J. Haidenbauer, M. Mai, R. Murray-Smith and D. R\"onchen, \href{https://doi.org/10.1103/PhysRevLett.123.182301}{Phys. Rev. Lett. {\bf 123}, 182301 (2019).}
\bibitem{Liu:2023xhg}
H.~Liu, J.~Zhang and X.~Wang,
\href{https://www.mdpi.com/2073-8994/15/1/214}{Symmetry \textbf{15}, 214 (2023).}

\bibitem{Lambda} M. Ablikim {\it et al.} (BESIII Collaboration), \href{https://doi.org/10.1038/s41567-019-0494-8}{Nature Phys. {\bf 15}, 631 (2019).}
\bibitem{3773Lambda} M. Ablikim {\it et al.} (BESIII Collaboration), \href{https://doi.org/10.1103/PhysRevD.105.L011101}{Phys. Rev. D {\bf 105}, L011101 (2022).}
\bibitem{zhangjianyu} M. Ablikim {\it et al.} (BESIII Collaboration), \href{https://doi.org/10.1103/PhysRevLett.129.131801}{Phys. Rev. Lett. {\bf 129}, 131801 (2022).}
\bibitem{Sigmap} M. Ablikim {\it et al.} (BESIII Collaboration), \href{https://doi.org/10.1103/PhysRevLett.125.052004}{Phys. Rev. Lett. {\bf 125}. 052004 (2020).}
\bibitem{Patrik} M. Ablikim  {\it et al.} (BESIII Collaboration), \href{https://doi.org/10.1038/s41586-022-04624-1}{Nature {\bf 606}, 64 (2022).}
\bibitem{zhangjingxu} M. Ablikim {\it et al.} (BESIII Collaboration), \href{https://doi.org/10.1103/PhysRevD.106.L091101}{Phys. Rev. D \textbf{106}, L091101 (2022).}
\bibitem{product1} R. Handler {\it et al.}, \href{https://doi.org/10.1103/PhysRevD.25.639}{Phys. Rev. D {\bf 25}, 639 (1982).}
\bibitem{product2} J. R. Batley {\it et al.} (NA48/1 Collaboration), \href{https://doi.org/10.1016/j.physletb.2010.08.046}{Phys. Lett. B {\bf 693}, 241 (2010).} 
\bibitem{HBC1} J. P. Berge, P. Eberhard, J. R.~Hubbard, D. W. Merrill, J. Button-Shafer, F. T. Solmitz and M. L. Stevenson, \href{https://doi.org/10.1103/PhysRev.147.945}{Phys. Rev. {\bf 147}, 945 (1966).}
\bibitem{HBC2} P. M. Dauber, J. P. Berge, J. R. Hubbard, D. W. Merrill and R. A. Muller, \href{https://doi.org/10.1103/PhysRev.179.1262}{Phys. Rev. {\bf 179} 1262 (1969).}
\bibitem{HBC3} C. Baltay, A. Bridgewater, W. A. Cooper, L. K. Gershwin, M. Habibi, M. Kalelkar, N. Yeh and A. Gaigalas, \href{https://doi.org/10.1103/PhysRevD.9.49}{Phys. Rev. D {\bf 9}, 49 (1974).}
\bibitem{LeeYang} T. D. Lee and C. N. Yang, \href{https://doi.org/10.1103/PhysRev.108.1645}{Phys. Rev. {\bf 108} 1645 (1957).}
\bibitem{formula} E. Perotti, G. F\"aldt, A. Kupsc, S. Leupold and
  J. J. Song,
  \href{https://doi.org/10.1103/PhysRevD.99.056008}{Phys. Rev. D {\bf
      99}, 056008 (2019).}
\bibitem{polarization} G. F\"aldt and A. Kupsc, \href{https://doi.org/10.1016/j.physletb.2017.06.011}{Phys. Lett. B \bf{772}, 16 (2017).}
\bibitem{psip} M. Ablikim {\it et al.} (BESIII Collaboration), \href{https://doi.org/10.1088/1674-1137/42/2/023001}{Chin. Phys. C {\bf 42}, 023001 (2018).}
\bibitem{BESIII} M. Ablikim {\it et al.} (BESIII Collaboration), \href{https://doi.org/10.1016/j.nima.2009.12.050}{Nucl. Instrum. Meth. A {\bf 614}, 345 (2010).}
\bibitem{TopoAna}	X. Zhou, S. Du, G. Li and C. Shen, \href{https://doi.org/10.1016/j.cpc.2020.107540}{Comput. Phy. Commun. {\bf 258}, 107540 (2021).}
\bibitem{kkmc1} S. Jadach, B. F. L. Ward and Z. Was, \href{https://doi.org/10.1016/S0010-4655(00)00048-5}{Comput. Phys. Commun.  {\bf 130}, 260 (2000).}
\bibitem{kkmc2} S. Jadach, B. F. L. Ward and Z. Was, \href{https://doi.org/10.1103/PhysRevD.63.113009}{Phys. Rev. D {\bf 63}, 113009  (2001).}
\bibitem{evtgen1} D. J. Lange, \href{https://doi.org/10.1016/S0168-9002(01)00089-4}{Nucl. Instrum. Meth. A {\bf 462}, 152 (2001).}
\bibitem{evtgen2} R. G. Ping, \href{https://doi.org/10.1088/1674-1137/32/8/001}{Chin. Phys. C {\bf 32}, 599 (2008).}
\bibitem{PDG2022}   R. L. Workma {\it et al.} (Particle Data Group), \href{https://doi.org/10.1103/PhysRevD.62.034003}{Prog. Theor. Exp. Phys. {\bf 2022}, 083C01 (2022).}
\bibitem{lundcharm1} J. C. Chen, G. S. Huang, X. R. Qi, D. H. Zhang and Y. S. Zhu, \href{https://doi.org/10.1103/PhysRevD.62.034003}{Phys. Rev. D {\bf 62}, 034003 (2000).}
\bibitem{lundcharm2} R. L. Yang, R. G. Ping and H. Chen, \href{https://doi.org/10.1088/0256-307X/31/6/061301}{Chin. Phys. Lett.  {\bf 31}, 061301 (2014).}
\bibitem{geant41}  S.Agostinelli {\it et al.} (GEANT4 Collaboration), \href{https://doi.org/10.1016/S0168-9002(03)01368-8}{Nucl. Instrum. Meth. A {\bf 506}, 250  (2003).}
\bibitem{geant42} J. Allison {\it et al.}, \href{https://doi.org/10.1109/TNS.2006.869826}{IEEE Trans. Nucl. Sci.  {\bf 53}, 270  (2006).}
\bibitem{BESIII:2016ssr}
M. Ablikim {\it et al.} (BESIII Collaboration), \href{https://journals.aps.org/prd/abstract/10.1103/PhysRevD.93.072003}{Phys. Rev. D \textbf{93}, 072003 (2016).}
\bibitem{preXi0} M. Ablikim {\it et al.} (BESIII Collaboration), \href{https://doi.org/10.1016/j.physletb.2017.04.048}{Phys. Lett. B {\bf 770}, 217 (2017).}
\bibitem{BESIII:2019dve}
M. Ablikim {\it et al.} (BESIII Collaboration), 
\href{https://journals.aps.org/prd/abstract/10.1103/PhysRevD.100.051101}{Phys. Rev. D \textbf{100}, 051101 (2019).}
\bibitem{BESIII:2019cuv}
M. Ablikim {\it et al.} (BESIII Collaboration), 
\href{https://journals.aps.org/prl/abstract/10.1103/PhysRevLett.124.032002}{Phys. Rev. Lett. \textbf{124}, 032002 (2020).}
\bibitem{BESIII:2020ktn}
M. Ablikim {\it et al.} (BESIII Collaboration), 
\href{https://journals.aps.org/prd/abstract/10.1103/PhysRevD.103.012005}{Phys. Rev. D \textbf{103}, 012005 (2021).}
\bibitem{BESIII:2021aer}
M. Ablikim {\it et al.} (BESIII Collaboration), 
\href{https://www.sciencedirect.com/science/article/pii/S0370269321004974?via%3Dihub}{Phys. Lett. B \textbf{820}, 136557 (2021).}
\bibitem{BESIII:2021ccp}
M. Ablikim {\it et al.} (BESIII Collaboration), 
\href{https://journals.aps.org/prd/abstract/10.1103/PhysRevD.104.L091104}{Phys. Rev. D \textbf{104}, L091104 (2021).}
\bibitem{BESIII:2021gca}
M. Ablikim {\it et al.} (BESIII Collaboration), 
\href{https://journals.aps.org/prd/abstract/10.1103/PhysRevD.104.092012}{Phys. Rev. D \textbf{104}, 092012 (2021).}
\bibitem{BESIII:2022mfx}
M. Ablikim {\it et al.} (BESIII Collaboration), 
\href{https://link.springer.com/article/10.1007/JHEP06(2022)074}{J. High Energy Phys. \textbf{06} (2022) 74.}
\bibitem{newdata} M. Ablikim {\it et al.} (BESIII Collaboration), \href{https://doi.org/10.1088/1674-1137/44/4/040001}{Chin. Phys. C {\bf 44}, 040001 (2020).}
\bibitem{panda} W. Erni {\it et al.} (PANDA Collaboration), \href{https://arxiv.org/abs/0903.3905}{arXiv 0903.3905 (2009) [hep-ex].}
\bibitem{superTC} A. E. Bondar {\it et al.} (Charm-Tau Factory Collaboration), \href{https://doi.org/10.1134/S1063778813090032}{Phys. Atom. Nucl. {\bf 76}, 1072 (2013).}
\bibitem{STCF} X. D. Shi, X. R. Zhou, X. S. Qin and H. P. Peng, \href{https://doi.org/10.1088/1748-0221/16/03/P03029}{J.~Instrum. {\bf 16}, P03029 (2021).}
\end{thebibliography}
\end{document}